\documentclass[sigconf]{acmart}
\usepackage{multirow}
\usepackage{wrapfig}
\usepackage{subfig}
\usepackage{cleveref}
\usepackage{algorithmicx}
\usepackage[Algorithm,ruled]{algorithm}

\AtBeginDocument{%
  \providecommand\BibTeX{{%
    \normalfont B\kern-0.5em{\scshape i\kern-0.25em b}\kern-0.8em\TeX}}}

\setcopyright{acmcopyright}
\copyrightyear{2018}
\acmYear{2018}
\acmDOI{10.1145/1122445.1122456}

\acmConference[Woodstock '18]{Woodstock '18: ACM Symposium on Neural
  Gaze Detection}{June 03--05, 2018}{Woodstock, NY}
\acmBooktitle{Woodstock '18: ACM Symposium on Neural Gaze Detection,
  June 03--05, 2018, Woodstock, NY}
\acmPrice{15.00}
\acmISBN{978-1-4503-XXXX-X/18/06}

\newcommand{\name}{{\sc Jigsaw}}
\newcommand{\sbpName}{{\sc Structured Partial Backpropagation}}
\newcommand{\sbpAcrnm}{{\sc SPB}} 
\newcommand{\minisection}[1]{\noindent{\bf #1.}\hspace{1mm}}

\copyrightyear{2021} 
\acmYear{2021} 
\setcopyright{acmcopyright}\acmConference[DistributedML'21]{2nd International Workshop on Distributed Machine Learning DistributedML 2021}{December 7, 2021}{Virtual Event, Germany}
\acmBooktitle{2nd International Workshop on Distributed Machine Learning DistributedML 2021 (DistributedML'21), December 7, 2021, Virtual Event, Germany}
\acmPrice{15.00}
\acmDOI{10.1145/3488659.3493778}
\acmISBN{978-1-4503-9134-4/21/12}

\begin{document}

\title{Doing More by Doing Less: How Structured Partial Backpropagation Improves Deep Learning Clusters}

\author{Adarsh Kumar}
\email{adrshkm@amazon.com}
\affiliation{%
  \institution{Amazon Alexa AI}
   \country{}
}

\author{Kausik Subramanian}
\affiliation{%
  \institution{University of Wisconsin, Madison}
   \country{}
}

\author{Shivaram Venkataraman}
\affiliation{%
  \institution{University of Wisconsin, Madison}
   \country{}
}
\author{Aditya Akella}
\affiliation{%
  \institution{University of Texas at Austin}
   \country{}
}








\renewcommand{\shortauthors}{Kumar Adarsh, et al.}

\begin{abstract}
  Many organizations employ compute clusters equipped with
  accelerators such as GPUs and TPUs for training deep learning
  models in a distributed fashion. Training is resource-intensive,
  consuming significant compute, memory, and network resources. Many
  prior works explore how to reduce training resource footprint
  without impacting quality, but their focus on a subset of the
  bottlenecks (typically only the network) limits their ability to
  improve overall cluster utilization. In this work, we
  exploit the unique characteristics of deep learning workloads to
  propose \sbpName{}(\sbpAcrnm{}), a
  technique that systematically controls the amount of
  backpropagation at individual workers in distributed
  training. This simultaneously reduces network bandwidth, compute
  utilization, and memory footprint while preserving model quality. To efficiently leverage the benefits of \sbpAcrnm{} at cluster level, we introduce \name{}, a \sbpAcrnm{} aware scheduler, which does scheduling at the iteration level for Deep Learning Training(DLT) jobs. We find
  that \name{} can improve large scale cluster efficiency by
  as high as 28\%.
\end{abstract}

\maketitle

\section{Introduction}
With the widespread use of deep learning models in a variety of
applications, many organizations are employing large clusters of machines equipped
with hardware accelerators for model training. These clusters
are expensive to build and are power hungry. They are also
oversubscribed with Deep Learning Training(DLT) jobs often having to wait in
queues for hours to days before getting
scheduled~\cite{philly,tiresias}. Thus, understanding how to improve the utilization
of such clusters is an important goal. In this paper, we
develop a novel technique for improving deep learning
cluster efficiency that is centered around the idea of having
\textbf{DLT jobs do {\em less}} work which can be systematically
leveraged to \textbf{run {\em more} training jobs in a highly efficient manner}.

Today, training of DNNs is typically done using algorithms like stochastic gradient descent (SGD),
an iterative algorithm where every iteration computes gradients by running a forward pass
followed by a backward pass.  To scale SGD across machines the most commonly used method is data parallel
distributed training, where gradients computed at each worker are aggregated after every iteration
of training.  SGD can take thousands of iterations to converge and thus training a deep learning
model can take days or weeks~\cite{philly}.


Many recent works~\cite{powersgd,lin2020deep,alistarh2017qsgd} have considered how 
to reduce the resources required by a given job with the primary focus
being algorithmic techniques that can approximate gradients exchanged between workers during
distributed training without affecting accuracy.
Most of these techniques aim to \textbf{solely} minimize network communication and examples include
quantizing gradients~\cite{alistarh2017qsgd}, dropping parts of
the gradient~\cite{lin2020deep} or using low-rank updates~\cite{powersgd, wang2018atomo}. 


Unfortunately, these techniques provide limited improvements for deep
learning clusters equipped with modern hardware that are used to train multiple models
simultaneously~\cite{gandiva,tiresias,philly}. This is primarily due to following
reasons: First, while existing approaches optimize communication, they
fail to reduce the total amount of compute resources or memory required
during training. For example, if a gradient dropping scheme only
preserves the top-20\% of the gradient values for communication, we still need to
compute the gradients for all the parameters during the backward phase, ending up using the \textbf{same} compute.
Thus, prior approaches simply shift bottlenecks from the network to
compute in large multi-tenant clusters. 
Further, while running in a multi-tenant cluster, the goals for an organization
are typically to improve overall cluster-wide utilization~\cite{gandiva} 
(e.g., reduce makespan given a set of jobs) and only reducing the network
resources required provides limited cluster-wide benefits. 



\begin{figure}[H]
  \includegraphics[width=0.8\linewidth]{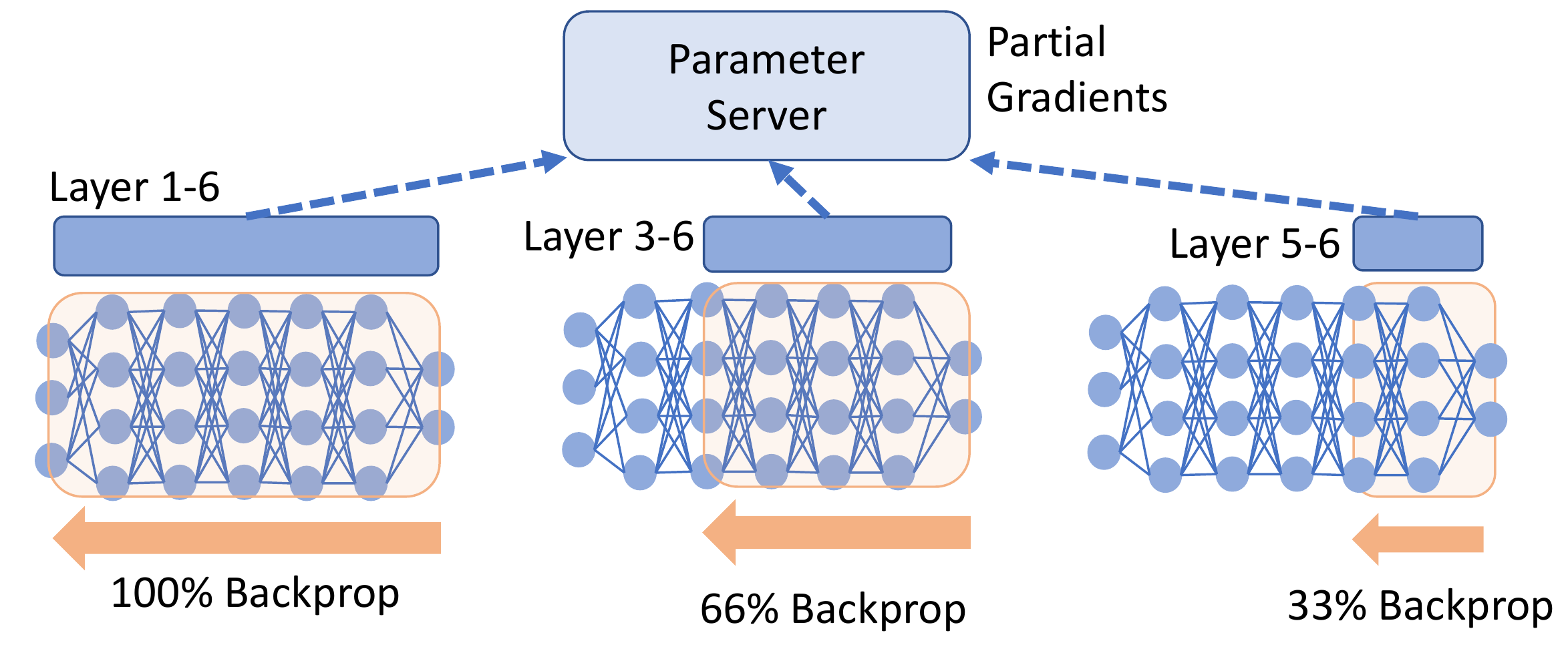}
  \vspace*{-5mm}
  \caption{Structured Partial Backpropagation (\sbpAcrnm)}
  \label{fig:sbp}
\end{figure}

This prompts us to ask the following question:

\begin{center}
\emph{``How can we design a \underline{distributed training scheme} that can
provide \underline{cluster-wide benefits} in a multi-tenant setting?''}
\end{center} 


To improve the overall cluster efficiency we need to improve the resource utilization of jobs across
both compute as well as communication dimension. 
Thus, techniques that avoid
computation of gradients which do not need to be communicated can lead to savings across both resources. 
Based on this insight, we design \sbpName{} (\sbpAcrnm{}), a new
gradient compression scheme that can provide cluster-wide benefits.
  

Our first observation in designing \sbpAcrnm{} is that
similar to existing schemes like top-K gradient dropping, 
we can approximate the gradient by updating
different parts of the overall gradient from different workers. For
example, consider a job which is spread across three workers as shown
in Figure~\ref{fig:sbp}. In this case, two of the three workers
compute partial gradients (33\% and 66\%) which are aggregated to
derive an overall approximate gradient. 
Assuming synchronous SGD, the above described scheme does not improve
the time taken by a single iteration as there is one worker which needs to compute the entire gradient.
However, this scheme does yield cluster-wide benefits as it
leads to an aggregate reduction in \textbf{network} utilization (similar to the
top-50\% scheme), while also saving \textbf{compute} resources along with \textbf{memory} as we describe next. 


The compute and memory savings in \sbpAcrnm{} arise from
the way gradients are computed for deep learning jobs. To compute a
gradient, first a forward pass is performed that computes the loss
value for the data examples with respect to the current
model. Following this, a backward pass is performed to compute the
final gradient values. When using \sbpAcrnm{}, we conduct the
backward pass for a fraction of the layers from the end\footnote{By ``end'' we refer to the final output layer of the DNN} of the DNN network, and this leads to reduced computation as
shown in Figure~\ref{fig:sbp}. The fraction of layers are carefully chosen to be the suffix layers as computing gradients for layer $k$ requires gradients from all layers $l$, where $l\geq k$.

\begin{table}
  \small
    \begin{minipage}{\columnwidth}
    \centering
      \begin{tabular}{p{4em} | p{4em} | p{4em} | p{4em}}
        \toprule
         \textbf{\% Backprop} & \textbf{Forward Pass (ms)} & \textbf{Backward Pass(ms) } & \textbf{Peak Mem (GB)} \\
        \midrule
        100 & 34.79 & 79.89  & 6.5 \\
        90 & 34.85 & 70.24 & 5.7\\
        80 & 34.71 & 58.86 & 5.1\\
        70 & 34.61 & 49.92 & 4.6 \\
        60 & 34.72 & 43.74 &  4.3\\
        50 & 34.70 & 36.58 & 4.1 \\
        40 & 34.61 & 29.93 & 3.8 \\
        30 & 34.67 & 23.02 & 3.5\\
        20 & 34.84 & 16.34 & 3.3\\
        10 & 34.86 & 8.35 & 3.1\\
        5 & 34.64 & 3.56 & 2.9\\
        \bottomrule
      \end{tabular}
      \caption{Effect of \% partial Backprop on Resource Utilization of ResNet101 model. BatchSize:64, Tesla V100}       
      \label{table:sbp_promise}
   \end{minipage}
   
\end{table}
\sbpAcrnm{} can also save memory, as we can avoid storing intermediate outputs for
layers that are not going to participate in the backward pass. This opens the door for space multiplexing in hardware accelerators.  


We theoretically prove that \sbpAcrnm{} converges in the same order as SGD and has bounded impact on model quality in a limited setting and empirically show minimal
impact on model quality across 10 models ranging from
computer vision models (e.g., ResNet-50, VGG-19 etc.) to NLP models
(CNN-Text, LSTM etc.).

While \sbpAcrnm{} can help reduce the resource requirement for
a single job, there is limited impact on cluster-wide resource
utilization without corresponding support from the job schedulers. 
Today, schedulers treat DLT jobs as fixed-work units and perform
gang scheduling such that every worker receives the same amount of
compute, memory and network resources. Thus, even if \sbpAcrnm{}
were used, existing
schedulers would not be able to capitalize on resources that become
available (e.g., $2^{nd}$/$3^{rd}$ worker in Fig~\ref{fig:sbp}).

To this end, we design a new cluster scheduler, \name{},
that can account for the variable amounts of computation that are
created as a result of \sbpAcrnm{} and pack computations from
different workloads at fine time-scales to improve
overall cluster utilization. Performing such packing requires reasoning about when and
how much resources are going to be available in the future. We perform
fine-grained iteration-level scheduling and exploit the repetitive
nature of DL workloads to derive a centralized scheduling policy that
intelligently places tasks to maximize resource
utilization. 

Our experiments on a 45-GPU cluster show that \name{}, in conjunction with \sbpAcrnm{}, can improve the makespan and cluster utilization by $~28\%$ while running a publicly available
trace of DL jobs~\cite{philly}. We also show that \name{} is more tolerant to resource contention in comparison to other state of the art DLT schedulers. 


\section{Structured Partial Backprop}
\label{sec:structured_pruning}
We now describe \sbpName{}, a distributed training scheme that can provide 
benefits across multiple resource dimensions. Consider a distributed training job running with $k$ workers, training a $L$ layer DNN. We assume
a parameter server (PS) setup~\cite{ps_osdi} for data parallel training where each 
worker computes the gradient for all $L$ layers and the gradient values are aggregated at the PS. The PS maintains a global copy of the model, which is updated after every iteration. 

We first describe the changes in the role of each worker with \sbpAcrnm{}. At
the start of an iteration $i$, as in existing frameworks, each worker fetches the
latest model version from the parameter server (PS). To make sure all
training data points contribute equally, each worker samples its batch
of data from the entire dataset and performs a forward pass to obtain
\begin{table}
  \caption{Median time and memory usage during forward and backward pass, and gradient size 
   of different DNN architectures on Tesla V100 (batch size of 128).}
	\small
		\centering
      \begin{tabular}{p{4em} | p{2.4em} | p{2.4em} | p{2.5em} | p{2.2em}| p{1em}}
        \toprule
        \multirow{2}{*}{\textbf{Model}} &
          \multicolumn{2}{c|}{\textbf{Forward}} &
          \multicolumn{2}{c|}{\textbf{Backward}} & \\
        & \textbf{Time (ms)} & \textbf{Mem (GB)} & \textbf{Time (ms)} & \textbf{Mem (GB)} &
        \textbf{Grad (MB)} \\
        \midrule
        ResNet18 & 9.19 & 0.05  & 21.49 & 2.46 & 44\\
        ResNet34 & 16.11 & 0.08 & 36.69 & 3.08 & 85  \\
        ResNet50 & 36.32 & 0.09 & 78.9 & 7.33 & 94  \\
        ResNet101 & 60.51 & 0.17 & 135.14 & 9.79 & 170 \\
        ResNet152 & 86.9 & 0.23 &  197.05 & 12.81 & 232 \\
        VGG19 & 6.82 & 0.08 & 16.31 & 2.02  &  80\\
        VGG16 & 5.68 & 0.06 & 13.96 & 1.97  & 59\\
        VGG11 & 3.34 & 0.04 & 7.8 & 1.83 & 36  \\
        GoogleNet & 41.33 & 0.05 & 99.17 & 5.96 & 24\\
        \bottomrule
      \end{tabular}
      \label{table:fwdbwdcomparison}
   \vspace{-5mm}
\end{table}
the loss with respect to the model. The main changes occur during the
backward pass: instead of performing the entire backward pass i.e. computing the gradient for all the layers, each
worker performs backpropagation on a fixed suffix of the network. As shown in
Figure~\ref{fig:sbp},
the $j^{th}$ worker does backward propagation for
$\frac{jL}{k}$ of the layers starting from the output layer of the
model. Finally, each worker sends the gradients computed to the
PS.

Similar to existing setups, the PS in \sbpAcrnm{} sends the latest version of
the model to all workers at the start of an iteration. The main
difference occurs in the aggregation phase: the PS now receives
partial gradient from each worker and needs to aggregate them to update
the model. The PS computes a weighted average, i.e., a layer's gradient
update is averaged by the effective number of workers that contributed to
it. Accordingly, the learning rate is also scaled with respect to
the number of workers contributing towards the gradient.

The above \emph{structured} approach to backpropagation has following notable aspects:

\noindent\textbf{Resource Savings}: As shown in Table~\ref{table:fwdbwdcomparison}, the backward pass
    uses more memory and takes longer to complete for DNNs. By doing backward pass for fraction of
    layers, \sbpAcrnm{} can save both compute and memory, thus enabling better time multiplexing of
    a GPU across jobs. Table:~\ref{table:sbp_promise} shows the savings when partial backpropagation is done for ResNet 101 model.

\noindent\textbf{Gradient Dependency}: Compute can only be saved with our
     proposed \sbpAcrnm{} approach. Avoiding backpropagation in Random-k or Top-k
     dropping is not possible because of the dependency of the
     gradients of early layers on later ones. 
     
\noindent\textbf{Per-Iteration Time}: On average, \sbpAcrnm{} reduces workers' running time.
     However, the per-iteration time remains
     the same as in synchronous SGD training as there's one
     worker that does complete backward propagation of all $L$ layers. Also, unlike previous gradient compression techniques like PowerSGD~\cite{powersgd} or Atomo~\cite{wang2018atomo}, we do not add any additional compute overhead to perform gradient compression.

\subsection{Theoretical Analysis}
We now present a worst-case theoretical analysis that proves that \sbpAcrnm{}
converges to an equivalent solution as traditional distributed SGD. We
approach this problem by first presenting the convergence rate for
traditional distributed SGD and then compare this to the convergence
rate when using \sbpAcrnm{}.

\noindent\textbf{Assumptions and Notation.} We denote $X$ as the
trainable parameters of the model and $f$ as the loss function. To
simplify our theoretical analysis we assume that $f:X \to\mathbb{R}$
is a convex function with $\beta$-smoothness\footnote{A continuously
  differentiable function $f$ is $\beta$-smooth if the gradient
  $\nabla f$ is $\beta$ Lipschitz continuous.}.

Given these assumptions the convergence rate for SGD is known from~\cite{bubeckoptimization} to be:
\begin{theorem}
  Let X be convex compact set and $R^2 = sup_{x, x_{1}\in X} ||x-x_{1}||^2$. If the
  approximate gradient $\tilde{g}(x)$ is such that $E||\nabla f(x) - \tilde{g}(x)||^2 \leq V^2$,  
  after $t$ iterations, SGD with step size $\frac{1}{\beta + \frac{1}{\eta(t)}}$,
  where $\eta(t) = \frac{R}{\sigma}\sqrt{\frac{2}{t}}$, achieves
    \begin{equation}
        E(f(\frac{1}{t}\sum_{s=1}^{t} x_{s+1}) - f(x^*)) \leq R\sqrt{\frac{2V^2}{t}} + \frac{\beta R^2}{t}
    \end{equation}
   \label{thm:sgd}
\end{theorem}

The above result bounds the difference between $f(\hat{x})$ after $t$ iterations and the optimal value
$f(x^*)$ in terms of $V^2$, which is a bound on how far our estimated gradient is from the true gradient.

Please see our Github repo\footnote{\url{https://github.com/adarsh-kr/Paper_JigSaw-}} for a complete proof.

\noindent\textbf{Distributed Mini-batch SGD.}
Now, we consider the scenario where we have $k$ workers each processing a batch $\frac{B}{k}$ at
every iteration. We first consider the scenario where every worker computes the entire gradient and
attempt to bound the value of $E||\nabla f(x) - \tilde{g}(x)||^2$.

\begin{lemma} 
  With $k$ workers, an overall batch size of $B$ and assuming
  $||g(x_i)|| < P$ $\forall i$, for mini-batch SGD, we have $E\|\nabla
  f(X) - \tilde{g}(x)\|^2 \leq 2P^2\frac{k}{B}$
\end{lemma}


\noindent\textbf{SPB Convergence Rate}
We next consider the case when \sbpAcrnm{} is used to prune gradients. The main difference here is that
$i^{th}$ block of the gradient vector will be updated by $i$ workers.

\begin{theorem} 
  When using \sbpAcrnm{} with $k$ workers and overall batch size of $B$ and assuming $||g(x_i)|| < P$ $\forall i$, 
  we have $E\|\nabla f(X) - \tilde{g}(x)\|^2 \leq 2P^2\frac{k}{B}log(k)$
\end{theorem}


The above result shows that when using \sbpAcrnm{} we can similarly bound the
value of $E||\nabla f(x) - \tilde{g}(x)||^2$, but the bound here has
an additional factor of $log(k)$ where $k$ is the number of workers.
Applying this result to Theorem~\ref{thm:sgd}, we see that the
convergence rate when using \sbpAcrnm{} is, in the worst case, $log(k)$
slower, i.e., when using \sbpAcrnm{} we might require a factor of $log(k)$
more iterations to converge to a similar quality result. However, this
is a worst-case bound for convex functions. We empirically show how
\sbpAcrnm{} performs in practice for real-world deep learning models next.

\begin{figure*}
  \centering
  \subfloat[Job DAG Structure]{\raisebox{4mm}{\includegraphics[width=0.24\linewidth]{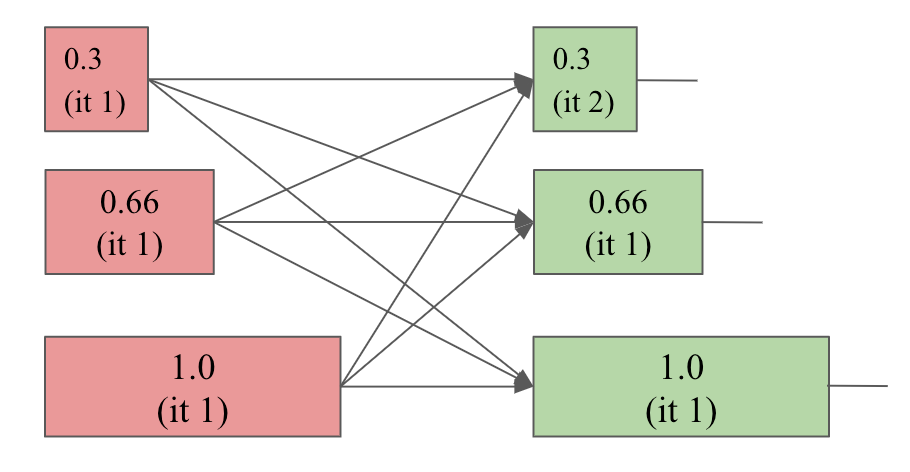}}}
  \hspace*{1mm}
  \subfloat[Gang Scheduling with \sbpAcrnm{}]{\includegraphics[width=0.37\linewidth]{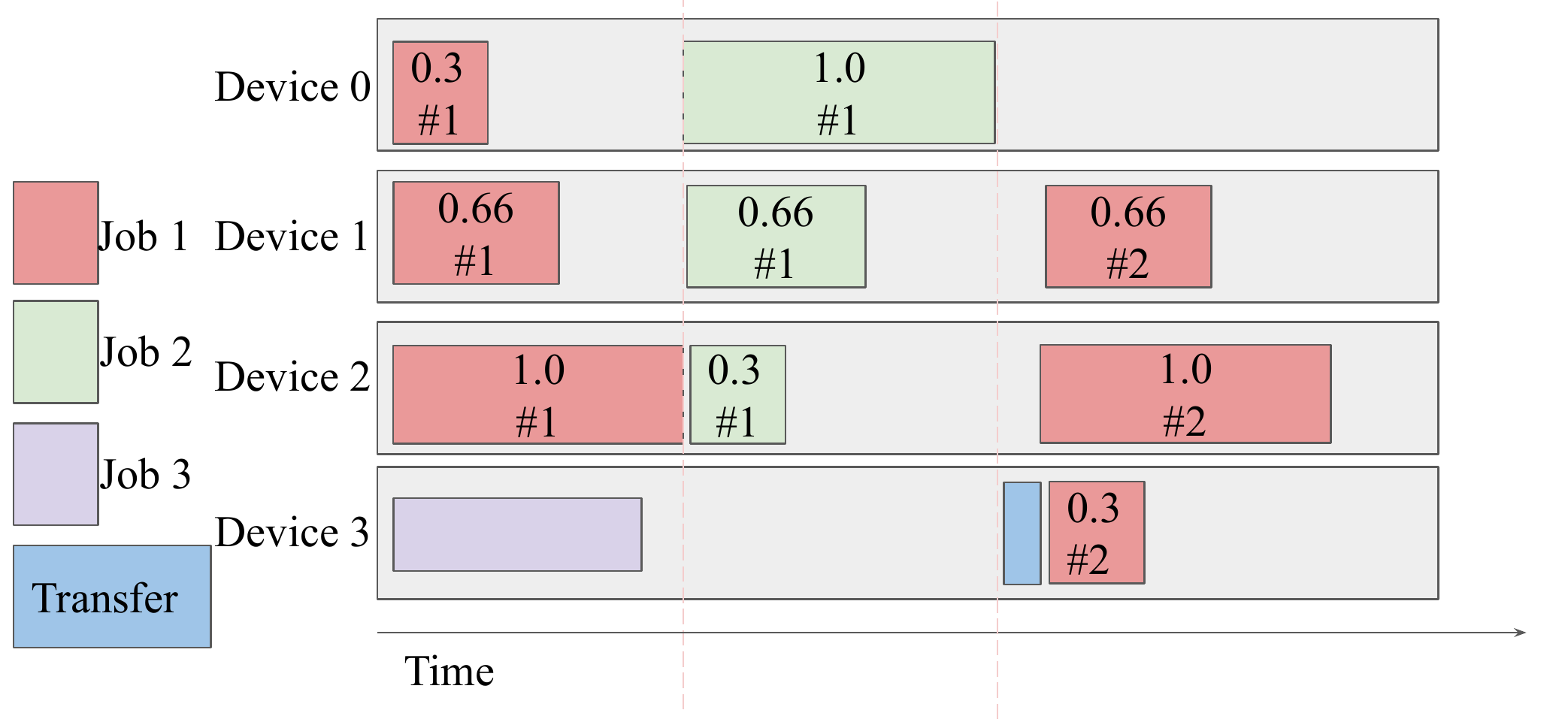}}
  \subfloat[Iteration-level scheduling with \sbpAcrnm{}]{\includegraphics[width=0.37\linewidth]{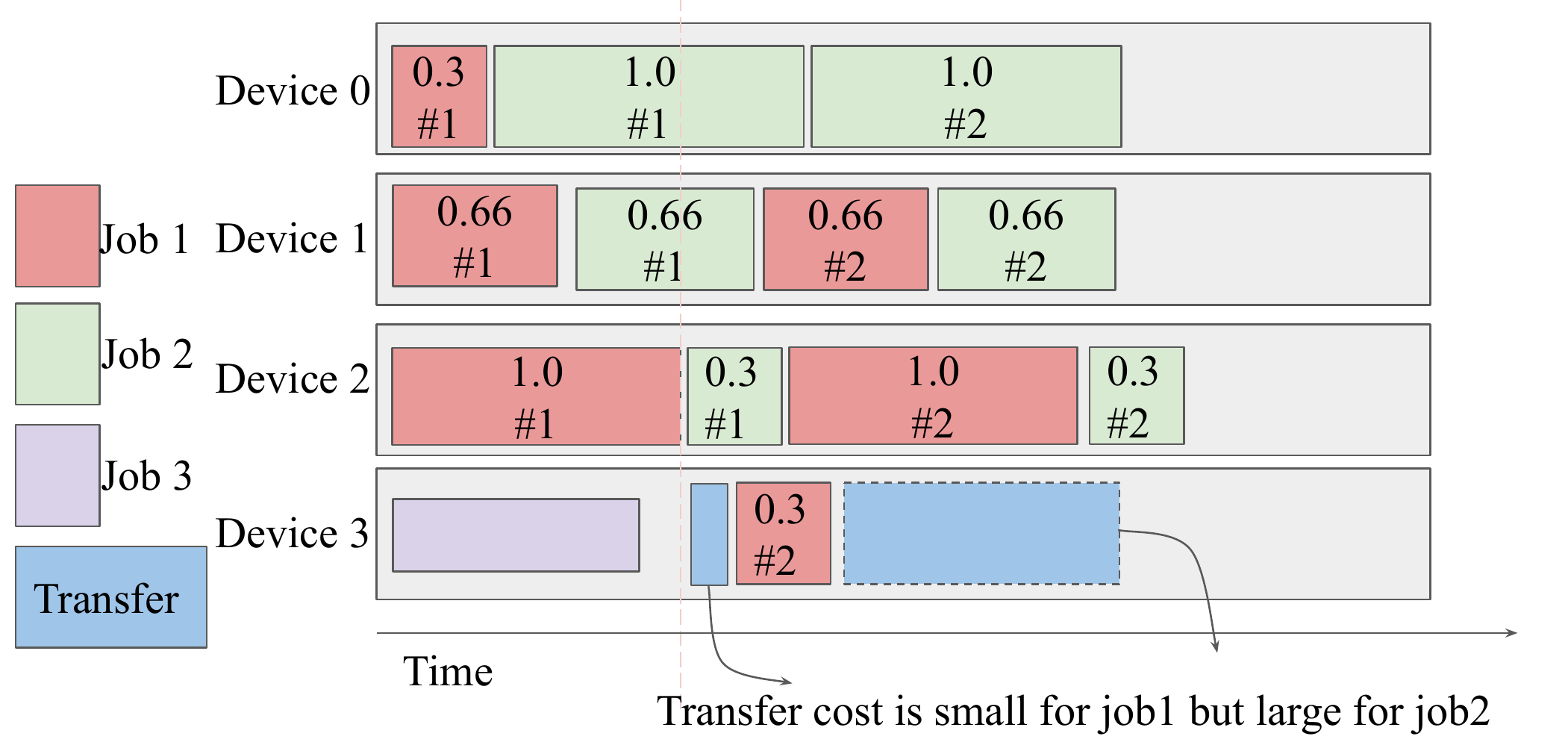}}
  \caption{\label{fig:job-schedule} 
  (a) DAG of a DNN training job trained with \sbpAcrnm{} with arrows indicating dependencies. 
  Tasks of iteration 2 can run 
  only after all tasks of iteration 1 have completed. Worker sizes indicate varying work distribution.
  Example of (b) gang scheduling versus (c) iteration-level 
  scheduling. Gang scheduling, which operates at a job-level, cannot utilize the resource savings generated
  by \sbpAcrnm{}.}
\vspace*{-4mm}
\end{figure*}

\section{\name{}: A \sbpAcrnm{} aware scheduler}
\label{sec:scheduling}

%

In this section, we describe the \name{} scheduler and how it is 
designed to leverage the gains obtained from our \sbpAcrnm{} to maximize cluster efficiency. We first describe some unique properties of \sbpAcrnm{} jobs that our scheduler needs to consider.
Following that we present our scheduling algorithm that aims to improve cluster
utilization and makespan.

\Cref{fig:job-schedule}(a) shows the task DAG of a single 
deep learning training job trained with \sbpAcrnm{} with three workers; for each iteration, we have multiple tasks running on parallel workers. Once the workers have computed 
the gradients for iteration $i$ and updated the parameter server, 
we can schedule tasks corresponding to iteration $i + 1$. Each 
task has two resources in its specification: peak GPU memory 
required, and running time. We note that these resource specifications
can be obtained by profiling a job for a few iterations~\cite{gandiva}.
Different tasks in the same iteration can have different resource 
characteristics due to \sbpAcrnm{}.

Given a set of DLT task DAGs, the \name{} scheduler computes
\emph{when} a task should run, and on \emph{which} machine.


\subsection{Iteration-Level Scheduling}
Existing state of the art DLT schedulers like Gandiva~\cite{gandiva} and
Tiresias~\cite{tiresias} perform scheduling at the
granularity of jobs, and rely on the paradigm of {\em gang scheduling}, i.e., for each
iteration, all parallel workers begin and end at the same time. However performing scheduling
only when all workers have finished one (or more) iterations is insufficient to make use
of resources saved by \sbpAcrnm{}.

Consider the example shown in Figure~\ref{fig:job-schedule}(b). In this case we have two parallel
DLT jobs (Job1, Job2) each of which has workers using \sbpAcrnm{}. Thus one of the workers finishes in 0.3s while the other two take 0.6s and 1s respectively. With gang scheduling we can see that the
scheduler is only invoked once all tasks of the iteration complete. Thus even though Device0 has
finished worker0 of Job1 at 0.3s, it remains idle for 0.7s before being assigned a new task.
By scheduling at the \emph{granularity of iterations}, we can utilize the gains from
\sbpAcrnm{} and utilize the cluster in a more effective way. In Figure~\ref{fig:job-schedule}(c), we
can see that each worker in a job can start at a staggered time (e.g. for Job2) but they still follow
the dependencies where one iteration is completed before the next one starts. In this case as
soon as
Device0 is done with worker0 of Job1 at 0.3s, it can be assigned to start Job2 thus
improving resource utilization and makespan for the cluster.

\noindent\textbf{Scheduling overheads.} However performing iteration-level scheduling can be
challenging as there is additional overhead for a centralized cluster scheduler to come up
with iteration-level schedules as iterations can be very short (100ms). 
To overcome the challenge of generating schedules efficiently, in \name{} we utilize the fact
that ML-training jobs are iterative and thus the scheduler has visibility into what tasks may arrive in
the future from a given job. Thus we perform coarse-grained scheduling (on the granularity of minutes) of fine-grained iterations~\cite{venkataraman2017drizzle}. 

\noindent\textbf{Task switching overheads.}
Switching between tasks at a fine granularity on a device can be expensive, especially
when workers are moved to a new device. This arises from the fact that moving a worker to a new
device entails fetching the model's computation graph and its latest parameters to the GPU 
before training can resume. We find that the model parameters and computation graph are
typically much smaller~\cite{salus} compared to the intermediate data generated.
Thus for workers that are time-sharing in the same device (e.g. Device1 in
Figure~\ref{fig:job-schedule}(b) we retain the model parameters in the global memory of the GPU and
this helps reduce switching overheads. For moving models across machines, as shown (e.g. Device3 in
Figure~\ref{fig:job-schedule}(b)), we account for the overhead of data transfer as a part of
scheduling. 

\vspace{-4mm}
\subsection{Scheduling Algorithm}
We next describe our scheduling algorithm. 
The scheduler computes the task placement for a scheduling 
interval $\mathcal{I}$. Our heuristic places
schedulable tasks (taking into account iteration dependencies) 
in a 2-dimensional space of resources (compute/device) and time (RT-Space) and 
tries to minimize makespan.

\minisection{Task Prioritization}
All tasks that are ready to run are placed in priority queue.  Similar to existing big-data 
schedulers that aim to improve packing~\cite{tetris, graphene}, we prioritize tasks
that require greater amount of resources. We calculate this priority as a  
normalized product of the task's GPU resource and running time. 
Given the dependency structure across iterations, placing the task with the highest 
resource requirement first also ensures that overall job progress does not get delayed.



\minisection{Task Placement}
Given a task and a partially occupied RT-Space, our scheduler next decides 
when a task must be run and on which machine. 
There are multiple considerations which decide placement of a task - first
and foremost, we must wait for tasks corresponding to the previous iteration
to finish before we can schedule the current task. Furthermore, 
the machine must have resources available for the entire duration
of the task. 

The scheduler first finds the earliest time the task can be scheduled to run based on the end times of the tasks of the previous
iterations $T_{i-1}$. However, not all machines can start the task immediately
after $T_{i-1}$. If a machine ran the task corresponding to the same worker and
the $i-1$ iteration, the task is immediately schedulable as the machine can
reuse the model computation graph from previous iteration. For the other machines, we add
a startup time which is dependent on the size of the model ($\gamma \times
model\_size$) to capture the delay in transferring the model to a new machine.
To reduce makespan, we greedily schedule the task
on a machine where the task can start the \textbf{earliest} of all machines. 
Since the scheduler chooses a machine where we can start
the task the earliest, it will naturally prefer machines which ran the previous
iteration tasks and avoid unnecessary task movement across iterations.

\section{Evaluation}
\label{sec:eval}
\subsection{\sbpAcrnm{ } effect on Model Quality}
We first present empirical results on the convergence of popular deep learning models when using
\sbpAcrnm{} while training. We consider models from a number of domains including computer vision,
NLP etc. and also consider models of different complexity (i.e., depth or number of layers). 

\begin{figure}[H]
  \includegraphics[width=0.8\linewidth]{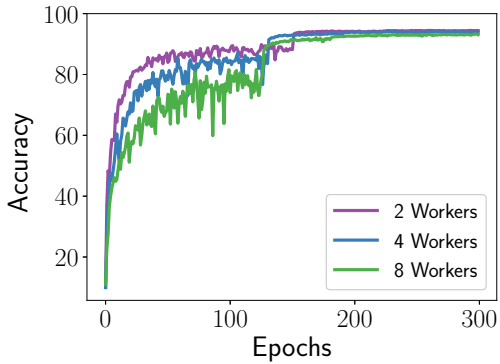}
  \vspace*{-5mm}
  \caption{Convergence of ResNet18 on CIFAR10 as we vary the number of workers. (\sbpAcrnm)}
  \label{fig:conv}
\end{figure}

For CV, we considered image classification
with CIFAR10 and CIFAR100 datasets and a number of DNN families
(ResNet, VGG, GoogLeNet, and DenseNet). We used SGD with momentum and
$10^{-4}$ weight decay as our optimizer configuration. We tried 3
learning rates for each model and picked the one attaining best
accuracy. For NLP, we selected models for classification on the MR
dataset~\cite{mr} and language modeling on
WikiText~\cite{wikitext}. We use 4 workers, each running on an NVIDIA
V100 GPU, for all the experiments.

 We present a comparison of the accuracy achieved when using
\sbpAcrnm{} vs. a baseline where models are trained using standard distributed SGD optimization in Table~\ref{table:sbp_accuracy}.

Our results from Table~\ref{table:sbp_accuracy} shows that there is a minimal effect on the accuracy achieved when models are trained with \sbpAcrnm{}. Even though the DNN early layers are getting updated by fewer workers, model convergence is not significantly impacted.
We attribute this behavior of \sbpAcrnm{} to the fact that early layers
converge faster than later layers, and thus require fewer gradient
descent steps. This has also been observed by previous
work~\cite{freezeOut,freezeTraining}. 

Further, 
we note that 
using \sbpAcrnm{} for a more complex task could lead to a bigger difference in
\begin{table}
  \small
  \centering
\begin{tabular}{c|c| c|c}
  \toprule
   \multirow{2}{*}{\textbf{Dataset}} &
    \multirow{2}{*}{\textbf{Model}} & \textbf{Accuracy} & \textbf{Accuracy}\\
    & & \textbf{\sbpAcrnm{}} & \textbf{SGD} \\
    \midrule
    \multirow{7}{*}{CIFAR10} & ResNet18 & 93.96 & 94.02 \\
    & ResNet34 & 93.81 & 94.96 \\
    & DenseNet121 & 94.18 & 94.35 \\
    & GoogleNet & 94.07 & 93.83 \\
    & VGG16 & 91.93 & 92.35 \\
    \hline
    \multirow{3}{*}{CIFAR100} & ResNet18 & 71.31 & 72.83 \\
    & ResNet34 & 69.94 & 71.2 \\
    \hline
    \multirow{2}{*}{MR} & CNNText & 77.12 & 77.35 \\
    & LSTMText & 78.02 & 80.23 \\
    \hline
    WikiText 2 & LSTM & 105* & 103* \\
    \bottomrule
    \end{tabular}
    \caption{Comparison of maximum accuracy of various models w/o SPB. *Perplexity is reported for language model on WikiText.} 
    \label{table:sbp_accuracy}
    \vspace*{-10mm}
  \end{table}
accuracy (though still under 2\%). For example, with ResNet18, the accuracy difference due to
\sbpAcrnm{} increases from 0.06\% to 1.52\% as we go from CIFAR10 to CIFAR100. This we believe can
be overcome by warm up epochs, where for initial $N$ epochs we train the model using the full
gradient from all the workers.

Figure~\ref{fig:conv} shows the convergence of ResNet18 model when varying the number of workers on the
CIFAR10 dataset while using SPB. We maintain a batch size of 128 per worker. 
From the figure we see that as we increase the number of workers it takes
more iterations to converge to the best accuracy. Our theory also suggests this in terms of the 
$log(k)$ factor in the convergence rate for $k$ workers, albeit for convex functions. 
Also, we find that the maximum accuracy achieved using 8 workers is slightly lesser than that 
achieved with 4 workers. We believe that this could be related to using a larger batch size in this
setting. We discuss ways in discussion section to close this gap in accuracy. 

\begin{figure*}
  \centering
\subfloat[Makespan comparison on 45 GPU Cluster]{\includegraphics[width=0.5\columnwidth]{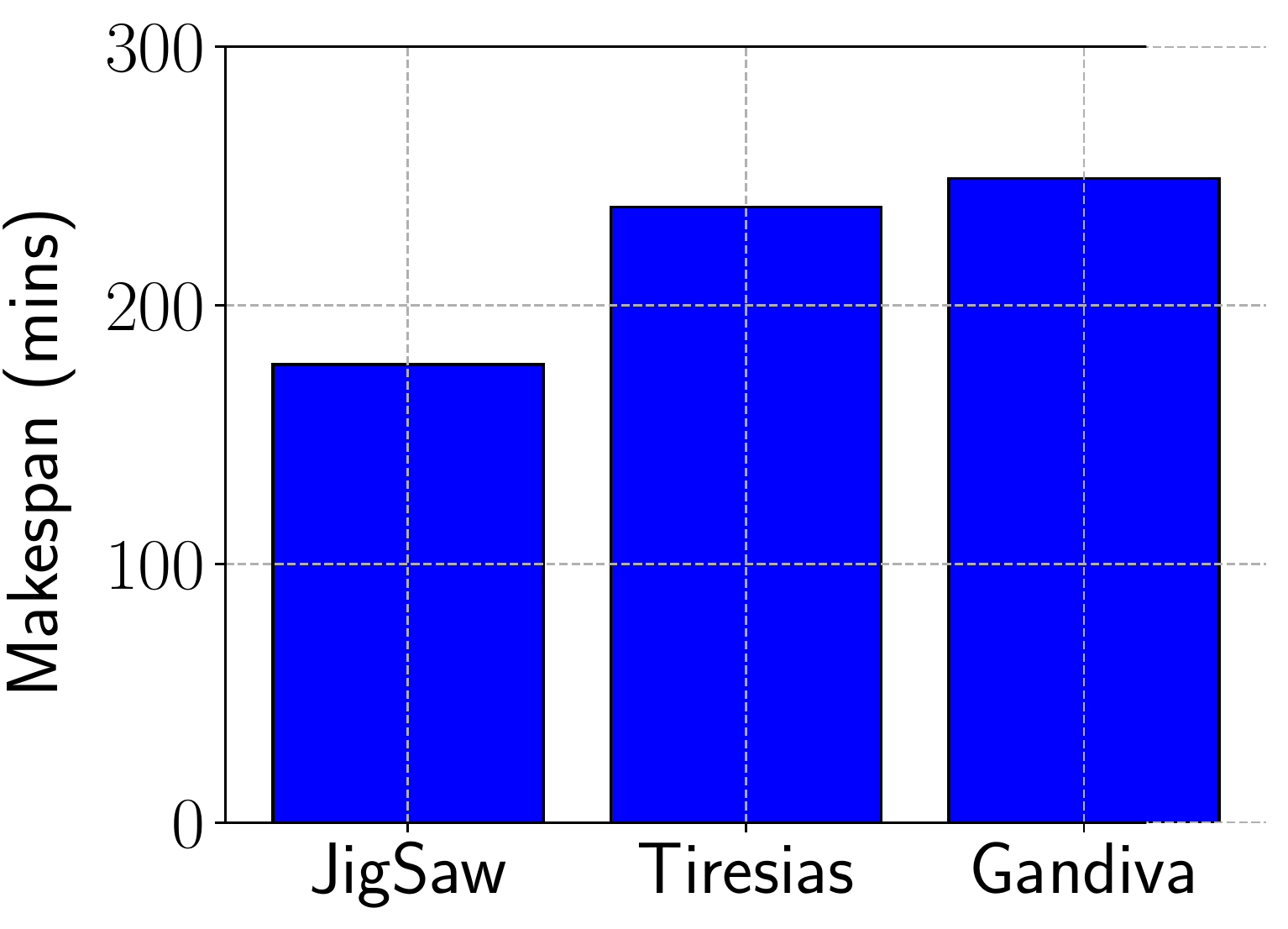}}\label{fig:makespan}
\subfloat[JCT comparison 45 GPU Cluster]{\includegraphics[width=0.5\columnwidth]{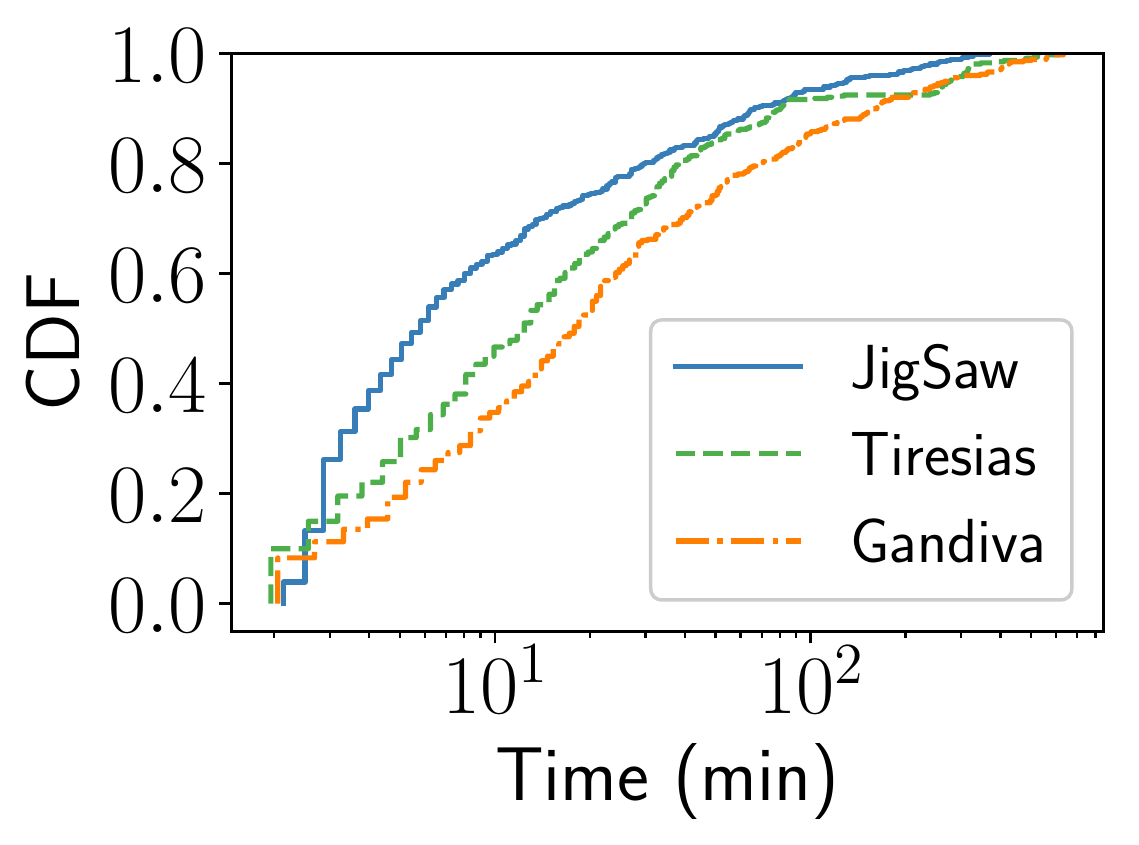}}
\subfloat[Worker Migrations]{\includegraphics[width=0.49\columnwidth]{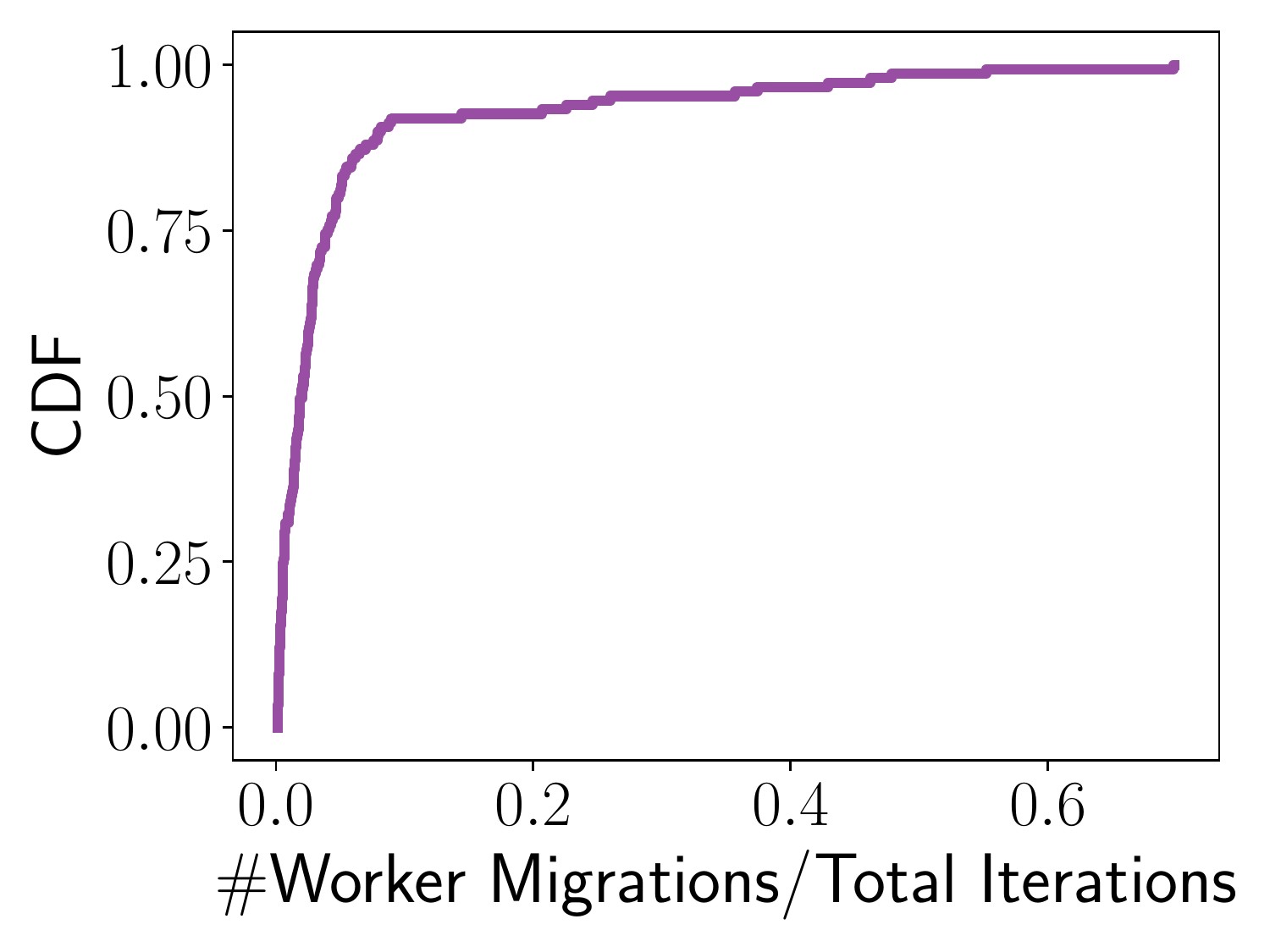}}
  \caption{\label{fig:workload_2}
       [Simulated]: (a) Makespan comparison on 45 Tesla V100 cluster. (b) CDF of JCT for different schedulers (c) CDF for fraction of iterations in which DLT jobs were migrated to another machine}
\end{figure*}

\subsection{\name{} effect on Cluster Efficiency}
In this section, we use a mix of simulations and cluster experiments to present 
the performance of our system \name{} and compare with other state of the art DLT schedulers.

\textbf{Setup:}
For our simulations, 
we created a discrete event based simulator which we run using a large scale DLT job trace from
Microsoft~\cite{tiresias,philly}. We enriched the trace with additional information, specifically
peak memory usage, execution time, gradient size and model size etc. by profiling models used
in~\citet{tiresias} trace. To enable \sbpAcrnm{}, we also profiled execution time of DLT workers
when varying the proportion of layers that perform backpropagation. 
All our profiling was done in isolation on Tesla V100 GPU. 
To account for migration overheads, we measure the transfer cost of moving models across different machines
and the network overhead in PS-worker communication. 

As described in Section~\ref{sec:scheduling}, \name{} creates a schedule for all the active DLT jobs for next $T$ minutes considering the complete cluster and communicates this
schedule to the individual machines. In our experiments we have kept $T$, the scheduling interval to
be 1 min. While creating the scheduling plan for next scheduling interval, \name{} considers the
affinity of tasks towards the machine they were executed previously to avoid
overheads from migration.            


\textbf{Workload}: We have evaluated \name{} on publicly available workload of DNN training from
Microsoft~\cite{tiresias} which was scaled down for practical purposes. This workload contains
around 500 DLT Training Job, with each job arriving at an average interval of 30 seconds. Around
50\% of the DLT jobs had 1 worker and the rest are multi-worker DLT jobs, with 10\% having 2, 20\% having 4, 15\% having 8 and 5\% having 16 workers. 
A DLT job is considered finished once a pre-specified number of iterations has been executed.

\textbf{Metrics}: To quantify and compare the performance of \name{}, we use two metrics:
\textbf{Job Completion Time(JCT):} JCT is defined as the time taken to finish one job from the time
it was submitted. This also includes any waiting time which the job had to incur. \textbf{Makespan:}
Makespan is the end time of the last job to finish in a trace. Improving makespan, correlates with improvement in resource utilization.

\textbf{Baselines}:
We evaluate our system \name{} against two state of the art DLT schedulers: Gandiva~\cite{gandiva},
which improves cluster efficiency by packing jobs on as few machines as possible and
Tiresias~\cite{tiresias}, which schedules job according to Least Attained Service(LAS) policy so
that all jobs obtain equal service over time. Both baselines are run standard DLT jobs, without
\sbpAcrnm{}, as their APIs do not allow workers of same DLT job to have different resource
requirements. 

\textbf{Results}:
Figure \ref{fig:workload_2} we compare makespan and JCT for Tiresias and Gandiva with \name{}. As can be seen, on 45 GPU Cluster, \name{} improves makespan by around 22\% over
Tiresias and 28\% over Gandiva.


Similarly, as can be seen from the CDF plot of JCT, our system \name{} improves the JCT of DLT jobs against Gandiva and Tiresias on a 45 GPU Cluster.

\vspace{-5mm}
\subsection{Testbed Overheads}
We discuss the systems overheads observed in our testbed. 

\textbf{Model Movements:} Migrating DLT job's worker across different machines is
costly~\cite{gandiva}. Since \name{} does fine grained scheduling at the level of
iteration, migrating the workers of DLT jobs at every iteration would result in significant
overheads. \name{} mitigates this by considering the migration overhead while deciding the placement
of a task. In Figure~\ref{fig:workload_2}(c) we plot the CDF of fraction of iterations for which a worker had
to be migrated to different machine. Fraction of iteration is calculated as total migrations divided by total iterations (lower the better). As can be seen from the plot, even after scheduling at iteration level \name{} has low migration overhead.               


\textbf{Time Multiplexing Overhead:} To improve the cluster utilization, \name{} also does time
multiplexing of DLT tasks. \sbpAcrnm{} creates opportunity for time multiplexing by finishing tasks
earlier by doing partial backprop. To study the overhead of time multiplexing of DLT tasks,
we ran 500 iterations of 5 real DNN models from Computer Vision and NLP family in sequence and in
a round robin manner. In sequence, the models are trained for 500 iterations one after the another,
while in round-robin we train a model for an iteration and then we switch to another model and
proceed in a round-robin manner. We found the 
slowdown to be less than 1\% during round-robin training, thus, the overhead of 
time multiplexing \textbf{within a GPU} is minimal.   


\section{Related Work}
\textbf{DL training schedulers}: There have been several recent works on improving cluster level
scheduling of DL training jobs, with each having a different objective
function~\cite{gandiva,tiresias,SLAQ,optimus}. For example, Gandiva~\cite{gandiva} focuses on improving
overall cluster efficiency, while Tiresias~\cite{tiresias} optimizes on job completion time and
fairness. Similar to Tiresias, Optimus~\cite{optimus} also focuses on average job completion time,
while SLAQ~\cite{SLAQ} optimizes on model quality.
All
existing DL training schedulers gang-schedule and assume equal resource requirement for all the
workers and hence cannot capitalize on resource savings from \sbpAcrnm{}.

\textbf{Gradient Pruning}: Ali et al~\cite{gradientDropping} introduced the idea of gradient
dropping. Strom et al~\cite{strom,ps_neurips} introduced dropping gradients larger
than a threshold and Adacomp~\cite{adacomp} proposed automatically tuning it. To further improve the convergence of
gradient dropping schemes, Chen et al~\cite{ba2016layer} introduced layer normalization. Other works
in this domain involve gradient quantization~\cite{seidi_1bit,signSGD,terngrad}, using low rank
estimation of gradients`\cite{powersgd}. All the above works focus on a single job setting where
compute is not a contended resource, and thus focus solely on improving communication. We introduce
a new structured pruning approach that can save compute, memory, and network resources.

\section{Conclusion}
We show how using a structured approach to partial backpropagation can improve 
resource utilization of DL training jobs in shared clusters.
We introduced a novel gradient pruning scheme called Structured
Backprop Pruning, which saves compute, memory and network resources, while
preserving model quality. To leverage the gains from \sbpAcrnm, we also
introduced a new SPB-aware cluster scheduler \name{} that can improve utilization by performing better 
time and space multiplexing. 
We show that our approach of making the cluster scheduler 
aware of the ML workload properties can lead to significant benefits.

\bibliographystyle{ACM-Reference-Format}
\bibliography{acmart.bib}
\include{appendix}
 \section{Appendix}

\subsection{Convergence Analysis}
We start by presenting the convergence theorem about SGD in ~\Cref{thm:sgd} from ~\cite{bubeckoptimization}

\begin{theorem}
    Let X be convex compact set and $R^2 = sup_{x, x_{1}\in X} ||x-x_{1}||^2$. If the
    approximate gradient $\tilde{g}(x)$ is such that $E||\nabla f(x) - \tilde{g}(x)||^2 \leq V^2$,  
    after $t$ iterations, SGD with step size $\frac{1}{\beta + \frac{1}{\eta(t)}}$,
    where $\eta(t) = \frac{R}{\sigma}\sqrt{\frac{2}{t}}$, achieves
      \begin{equation}
          E(f(\frac{1}{t}\sum_{s=1}^{t} x_{s+1}) - f(x^*)) \leq R\sqrt{\frac{2V^2}{t}} + \frac{\beta R^2}{t}
      \end{equation}
     \label{thm:sgd}
  \end{theorem}
  
  The above result bounds the difference between $f(\hat{x})$ after $t$ iterations and the optimal value
  $f(x^*)$ in terms of $V^2$, which is a bound on how far our estimated gradient is from the true gradient.
  
  \noindent\textbf{Distributed Mini-batch SGD.}
  Now, we consider the scenario where we have $k$ workers each processing a batch $\frac{B}{k}$ at
  every iteration where $B$ is cumulative mini batch size. We first consider the scenario where every worker computes the entire gradient and
  attempt to bound the value of $E||\nabla f(x) - \tilde{g}(x)||^2$.
  
  \begin{lemma} 
    With $k$ workers, an overall batch size of $B$ and assuming
    $||g(x_i)|| < P$ $\forall i$, for mini-batch SGD, we have $E\|\nabla
    f(X) - \tilde{g}(x)\|^2 \leq 2P^2\frac{k}{B}$
  \end{lemma}
  
  \textbf{Proof}: 
  We partition the parameter space $X$ into subparts, $x_i$
  \begin{equation}
    \tilde{g}(x,\epsilon) = [ ...., \frac{1}{B}\sum_{j=1}^{B} g_i(x_i,\epsilon_j), .... ]    
    \end{equation}
    
    For $i^th$ block of parameters, 
    
    \begin{equation} \label{eq1}
    \begin{split}
    E\|\nabla f(x_i) - \tilde{g}(x_i, \epsilon)\|^2 &= E\|\nabla f(x_i) - \frac{1}{B}\sum_{j=1}^{B}g_j(x_i)\|^2 \\
     & = \frac{1}{B}E\| \nabla f(x_i) - g_1(x_i)\|^2 \\
     & = \frac{1}{B}p_i
    \end{split}
    \end{equation}
    where we denote $ p_i$ as $E\| \nabla f(x_i) - g_1(x_i)\|^2$ \\
  
    For the overall parameter space this reduces to:
    \begin{equation}
        E\|\nabla f(x) - \tilde{g}(x, \epsilon)\|^2 = \sum_{i=1}^{k} \frac{1}{B}p_i
    \end{equation}

    Under the assumption that $||g(x_i)|| < P$   $\forall i$, we get 
    \begin{equation}
        E\|\nabla f(x) - \tilde{g}(x, \epsilon)\|^2 \leq 2P^2\frac{k}{B}
    \end{equation}

We now obtain the same bound for SPB case.\\

\noindent\textbf{SPB Convergence Rate}
We next consider the case when SPB is used to prune gradients. The main difference here is that
$i^{th}$ block of the gradient vector will be updated by $i$ workers.

\begin{lemma} 
  When using SPB with $k$ workers and overall batch size of $B$ and assuming $||g(x_i)|| < P$ $\forall i$, 
  we have $E\|\nabla f(X) - \tilde{g}(x)\|^2 \leq 2P^2\frac{k}{B}log(k)$
\end{lemma}
In SPB setting, $i^{th}$ block will be updated by i workers, leading to $\frac{iB}{k}$ mini batch, as each worker is processing $\frac{B}{k}$ mini batch.
\begin{equation}
\tilde{g_{SPB}(x,\epsilon)}) = [ ...., \frac{1}{\frac{iB}{k}}\sum_{j=1}^{\frac{iB}{k}} g_i(x_i,\epsilon_j), .... ] \\
\end{equation}
For the $i^{th}$ block of parameters, 

\begin{equation} \label{eq2}
\begin{split}
E\|\nabla f(x_i) - \tilde{g}(x_i, \epsilon)\|^2 &=  \frac{k}{iB}p_i
\end{split}
\end{equation}
Similar to previous baseline case, for the overall parameter space we get, 
\begin{equation} \label{eq3}
\begin{split}
E\|\nabla f(x) - \tilde{g}(x, \epsilon)\|^2 &= \sum_{i=1}^{k} \frac{k}{iB}p_i \\
 & \leq P^2\frac{k}{B}\sum_{i=1}^{k} \frac{1}{i} \\
 & \leq 2P^2\frac{k}{B}log(k) \\ 
\end{split}
\end{equation}

Substituting \Cref{eq3} and \Cref{eq2} in \Cref{thm:sgd}, we compare the convergence rate, which shows that SPB needs $log(k)$ times more iterations to achieve similar performance as the baseline setting where all workers are sending complete gradients. This bound suggests that on increasing the number of workers, logarithmically more iteration are going to be needed, which seems intuitive as the more the number of workers, the more fraction of data points that are not getting utilized in updating the parameters in comparison to baseline.

\end{document}